\documentclass[a4paper,fleqn,usenatbib]{mnras}

\usepackage{txfonts}

\usepackage[T1]{fontenc}
\usepackage{ae,aecompl}

\usepackage{algorithm}
\usepackage{algorithmic}
\usepackage{graphicx}	
\usepackage{booktabs} 
\usepackage{subcaption} 

\graphicspath{{images/}}

\title[AOLI TP3-WFS]{Laboratory and telescope demonstration of the TP3-WFS for the adaptive optics segment of AOLI}

\author[C. Colodro-Conde et al.]
{C. Colodro-Conde,$^{1}$\thanks{E-mail: carlos.colodro@upct.es}
S. Velasco,$^{2,3}$
J.J.F. Valdivia,$^{4,5}$
R.L. L\'opez,$^{2,3}$
A. Oscoz,$^{2,3}$
\newauthor
R. Rebolo,$^{2,3,6}$
B. Femen\'ia,$^{7}$
D.L. King,$^{8}$ 
L. Labadie,$^{9}$
C. Mackay,$^{8}$ \newauthor
B. Muthusubramanian,$^{9}$
A. P\'erez Garrido,$^{10}$
M. Puga,$^{2,3}$ 
G. Rodr\'iguez-Coira$^{2,3}$
\newauthor
L.F. Rodr\'iguez-Ramos,$^{2,3}$ 
J.M. Rodr\'iguez-Ramos,$^{4,5,11}$ 
R. Toledo-Moreo,$^{1}$
and
\newauthor
I. Vill\'o-P\'erez$^{1}$
\\ 
$^{1}$Departamento de Electr\'onica y Tecnolog\'ia de Computadoras, Universidad Polit\'ecnica de Cartagena, E-30202 Cartagena, Spain\\
$^{2}$Instituto de Astrof\'isica de Canarias, c/V\'ia L\'actea s/n, La Laguna, E-38205, Spain\\
$^{3}$Departamento de Astrof\'isica, Universidad de La Laguna, La Laguna, E-38200, Spain\\
$^{4}$Departamento de Ingenieria Industrial, Universidad de La Laguna, La Laguna, Spain.\\
$^{5}$Wooptix S.L., Torre Agust\'in Ar\'evalo, Avenida Trinidad, La Laguna, E-38205, Spain\\
$^{6}$Consejo Superior de Investigaciones Cient\'ificas, Madrid, Spain\\
$^{7}$W. M. Keck Observatory, 65-1120 Mamalahoa Hwy., Kamuela, HI 96743, Hawaii, USA\\
$^{8}$Institute of Astronomy, University of Cambridge, Madingley Road, Cambridge CB3 0HA, United Kingdom\\
$^{9}$I. Physikalsiches Institut, Universit\"{a}t zu K\"{o}ln, Z\"{u}lpicher Strasse 77, 50937 K\"{o}ln, Germany\\
$^{10}$Departamento de F\'isica Aplicada, Universidad Polit\'ecnica de Cartagena, Cartagena, E-30202, Spain\\
$^{11}$Centro de Investigaciones Biom\'edicas de Canarias, Campus Ciencias de La Salud s/n, E-38071 La Laguna, Spain}

\date{Accepted 2017 January 27. Received 2017 January 17; in original form 2016 October 18}

\pubyear{2017}

\begin{document}
\label{firstpage}
\pagerange{\pageref{firstpage}--\pageref{lastpage}}
\maketitle

\begin{abstract}
AOLI (Adaptive Optics Lucky Imager) is a state-of-art instrument that combines adaptive optics (AO) and lucky imaging (LI) with the objective of obtaining diffraction limited images in visible wavelength at mid- and big-size ground-based telescopes. The key innovation of AOLI is the development and use of the new TP3-WFS (Two Pupil Plane Positions Wavefront Sensor). The TP3-WFS, working in visible band, represents an advance over classical wavefront sensors such as the Shack-Hartmann WFS (SH-WFS) because it can theoretically use fainter natural reference stars, which would ultimately provide better sky coverages to AO instruments using this newer sensor. This paper describes the software, algorithms and procedures that enabled AOLI to become the first astronomical instrument performing real-time adaptive optics corrections in a telescope with this new type of WFS, including the first control-related results at the William Herschel Telescope (WHT).

\end{abstract}




\section{Introduction}

Reaching the diffraction limit in the visible wavelength is one of the main reasons to place optical telescopes on-board satellites such as the Hubble Space Telescope (HST), thereby avoiding the distortions and blurring that the atmosphere introduces on the unaltered wavefronts. With its life-cycle coming to end and without any oncoming substitute in visible bands, it is crucial to provide the scientific community with tools capable of providing similar resolutions from ground. In the era of the extremely large telescopes, and due to the increase of the atmospheric distortion as the diameter of the aperture grows, this has become a world top engineering challenge.

There are two main techniques which lead to diffraction-limited imaging. On the one side, Lucky Imaging (LI) \citep{1964JOSA...54...52H,1978JOSA...68.1651F, Brandner2016} offers an excellent and cheap method for reaching diffraction limited spatial resolution in the visible band in small and mid-size ground-based telescopes \citep{2008SPIE.7014E..47O}. However, this technique suffers from two important limitations. First, resolutions similar to the HST can only be achieved in telescopes with sizes below 2.5m \citep{2011MNRAS.413.1524F,2011A&A...526A.144L}. Second, most of the images are discarded, meaning that only relatively bright targets can be observed.

The other technique, Adaptive Optics (AO), has been the main procedure to improve the quality of the largest ground-based telescopes during the last twenty years \citep{1990A&A...230L..29R,1993ARA&A..31...13B,Milli2016}. The use of AO systems in infrared observations provides an adequate performance due to the reduced effects of turbulence in this wavelength range, thus achieving excellent results \citep{2010SPIE.7736E..24G, 2014PNAS..11112661M, 2014ApJ...791...35L}. Unfortunately, the scarce number of AO systems developed for the visible bands \citep{2008SPIE.7014E..18B, 2012SPIE.8447E..0XC} have not yet achieved the versatility and image quality already achieved in NIR \citep{2013ApJ...774...94C,2005ApJ...629..592G}, except for solar telescopes \citep{2012AN....333..863B}. The difficulty of performing AO in the visible wavelenghts is explained by the fact that the correlation time of the atmospheric turbulence scales with $\lambda^{6/5}$ \citep{1977JOSA...67..390G,1990JOSAA...7.1224F}, which means that an AO control loop operating in visible bands needs to be faster that in the NIR bands in order to provide the same degree of correction.

The Adaptive Optics Lucky Imager (AOLI) is a state-of-the-art instrument which was conceived to obtain extremely high resolution at optical wavelengths on big-sized telescopes by combining the two techniques presented in the paragraphs above (AO + LI) \citep{2015hsa8.conf..850V, 2016MNRAS.460.3519V}. Initially targeted for the 4.2m William Herschel Telescope (WHT, Observatorio del Roque de los Muchachos, Spain), the instrument is designed as a double system that encompasses an adaptive optics control system before the science part of the instrument, this last implementing LI \citep{2016arXiv160803230M}. Each of the two parts of AOLI can behave as a standalone system, meaning that the AO subsystem might be used with any other science instrument, performing imaging, spectroscopy or even coronography or polarimetry.

Besides the fact of combining AO and LI for the first time in an astronomical instrument, the other key innovation of AOLI 
is the development and use of a new type of wavefront sensor (WFS) in its AO subsystem: the Two Pupil Plane Positions Wavefront Sensor (TP3-WFS). The implementation of the TP3-WFS was motivated by the will of being able to use fainter AO reference stars than the ones that classical wavefront sensors such as the Shack-Hartmann WFS (SH-WFS) can use, all with the aim of increasing the sky coverage of the instrument without the need of laser guide stars.

The present work gives a comprehensive description of the AO subsystem of AOLI, focusing on the TP3-WFS and all the related algorithms and procedures that were developed for its characterization and testing. The first control-related results obtained with AOLI at the WHT, also included in the paper, confirm the viability of the TP3-WFS as part of a fully-functional adaptive optics system.

The paper is organized as follows. Section~\ref{sec:aoli_ao} introduces the AO subsystem of AOLI. Section~\ref{sec:arch} describes the AO real-time processing pipeline. Section~\ref{sec:sys_char} describes the procedures that were used to get information the AO subsystem, thus enabling the configuration the control algorithm. Section~\ref{sec:results} presents the results obtained both in laboratory and telescope tests. Finally, Section~\ref{sec:conclusions} draws the main conclusions.

\section{The Adaptive Optics System of AOLI}
\label{sec:aoli_ao}

\begin{figure*}
\centering
\includegraphics[scale=0.5]{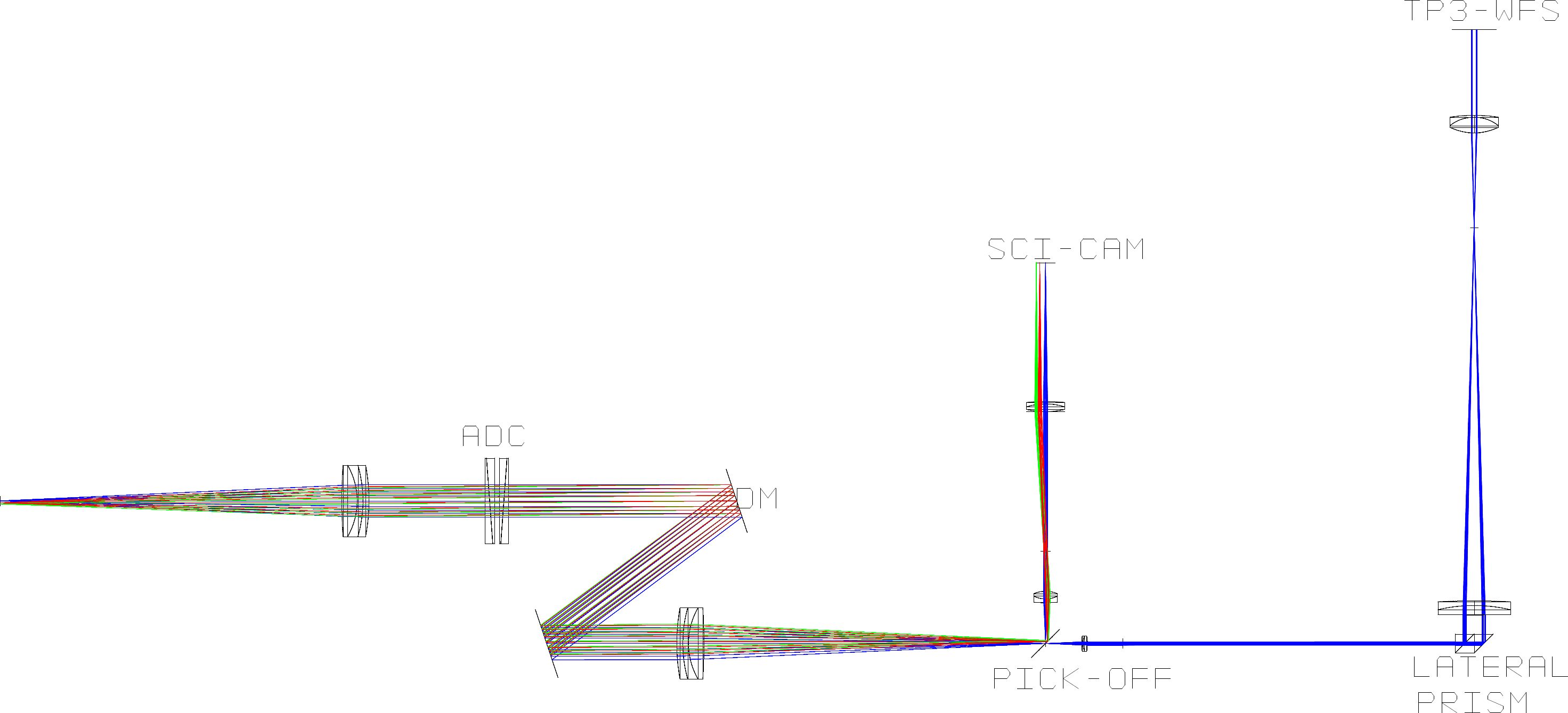}
\caption {AOLI optical layout. The system is divided in three modules: Deformable Mirror (DM), WaveFront Sensing (WFS) and SCIence (SCI). The common vertex is a Pick-off mirror that selects the reference star for wavefront sensing through a pin-hole in the mirror. This pin-hole can be selected as a real hole or with a splitting ratio $R/T=30/70$ and several sizes depending on the current seeing. The science arm can select two possible collimators to obtain two different Fields Of View (FOV). The WFS arm uses a lateral prism to introduce a differential delay between the two optical paths, using a common lens system to defocus the pupil image over the detector. The DM module is a typical 1:1 system with an Amici-biprims Atmospheric Dispersion Corrector (ADC).
}
\label{fig:aoli_zemax} 
\end{figure*}

\begin{figure*}
\centering
\includegraphics[scale=0.5]{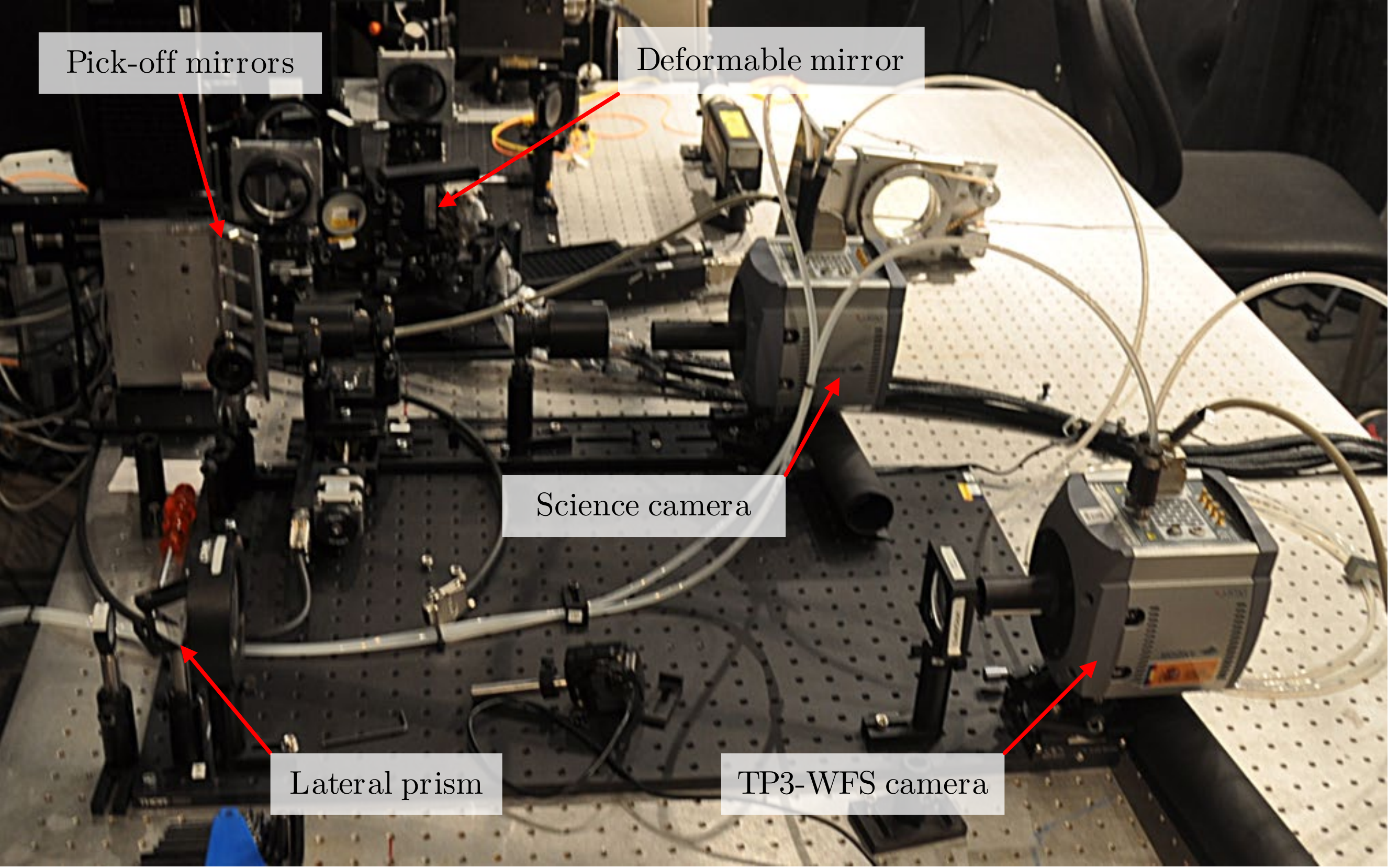}
\caption {AOLI mounted at one the two Nasmyth focuses of the WHT, inside the GRACE (GRound based Adaptive optics Controlled Environment) structure. The main components have been identified with labels.
}
\label{fig:aoli_photo} 
\end{figure*}

AOLI has been built putting together the expertise of several institutions, each group specialized in a different subject. To face the challenge that AOLI represents we have implemented a new philosophy of instrumental prototyping by modularizing all its components \citep{2016arXiv160804806L}: simulator/calibrator \citep{2014SPIE.9147E..7VP}, science module (performing LI) and AO module. Figure~\ref{fig:aoli_zemax} depicts the optical layout of AOLI, along with a description of the setup. On the other hand, Figure~\ref{fig:aoli_photo} shows a photograph of the instrument as mounted on the WHT.

The AO subsystem of AOLI comprises a 241-actuators deformable mirror (DM) by ALPAO, a pick-off guide-star subsystem and the Two Pupil Plane Positions Wavefront Sensor (TP3-WFS). The TP3-WFS operates with the images provided by an Andor Ixon DU-897 camera, which is based on a sub-photon noise 512x512 e2v EMCCD (Electron Multiplying Charge-Coupled Device) detector. The heart of the AO system is its Real-Time Control Software (RTC), which allows the control of 153 Zernike modes with a delay under 40$\mu$s. The delay of the calculations performed in the TP3-WFS itself is around 1 ms for the same number of reconstructed modes.

AOLI was required to perform wavefront sensing using faint reference stars up to magnitude 16 in the \textit{I} band with a seeing of 1 arcsec and at a wind speed of 8 km/s, which corresponds to the median value of the wind speed at the Roque de los Muchachos observatory. For the WHT this means sensing with up to four magnitudes fainter stars than the limit reached with a classical Shack-Hartmann WFS (SH-WFS). On the other hand, it was estimated that performing low-order AO corrections at a rate of 100 Hz in combination with LI would provide the desired level of correction at the science camera.

\subsection{The TP3-WFS}

The leading position for sensing the wavefront on AO systems has been occupied so far by the Shack-Hartmann WFS (SH-WFS), \citep{hartmann1900,shack1971production,platt2001history}. One important disadvantage of the SH-WFS is that the incoming photons are distributed among all the illuminated lenslets. This sets a limitation for the magnitude of the reference stars whose wavefronts can be reconstructed, and consequently to the sky coverage of an instrument based on this type of sensor. We have developed the TP3-WFS with the aim of overcoming this disadvantage.

The TP3-WFS bases its calculations on the intensity of the images of two defocused pupil images taken at two different planes. Computer simulations predict that this way of measuring wavefronts allows attaining good reconstructions with down to 100 photons falling within each pupil image \citep{vanDam:02}. Although this statement is yet to be thoroughly tested under a real environment (not only by simulations), if confirmed it would mean a considerable improvement in the sensitivity when compared to the SH-WFS. Another advantage of the TP3-WFS with respect of previous wavefront sensors such as the SH-WFS is that it is capable of working on extended targets, as demonstrated later in Section~\ref{sec:results}.

The TP3-WFS is composed of the wavefront reconstruction software (WFR), the WFS camera and the surrounding optics (see Figure~\ref{fig:aoli_zemax}). The main function of the TP3-WFS optics is to obtain the two defocused images near the pupil plane, which are later acquired by the WFS camera and finally processed by the WFR software. The WFR software is founded on the algorithm proposed by \cite{vanDam:02}, and it operates in real-time thanks to its GPU-accelerated implementation \citep{2013OptEn..52e6601F}. More detail about the internals of the WFR software will be given later in Section~\ref{sec:wfr}.


\section{AO processing pipeline}
\label{sec:arch}

Had the AO system of AOLI been based on well-known technologies such as the Shack-Hartmann WFS, the team could have benefited from the developments and knowledge from previous projects, or even from commercially available solutions. However, the use of a type of WFS that had never been implemented before for astronomical applications forced the development of not only the WFS itself, but also of the surrounding methods and software. The new system had to be designed for the specific needs of the TP3-WFS, with a focus on flexibility so as to counter uncertainties that this new method arose.

\begin{figure*}
\centering
\includegraphics[scale=0.8]{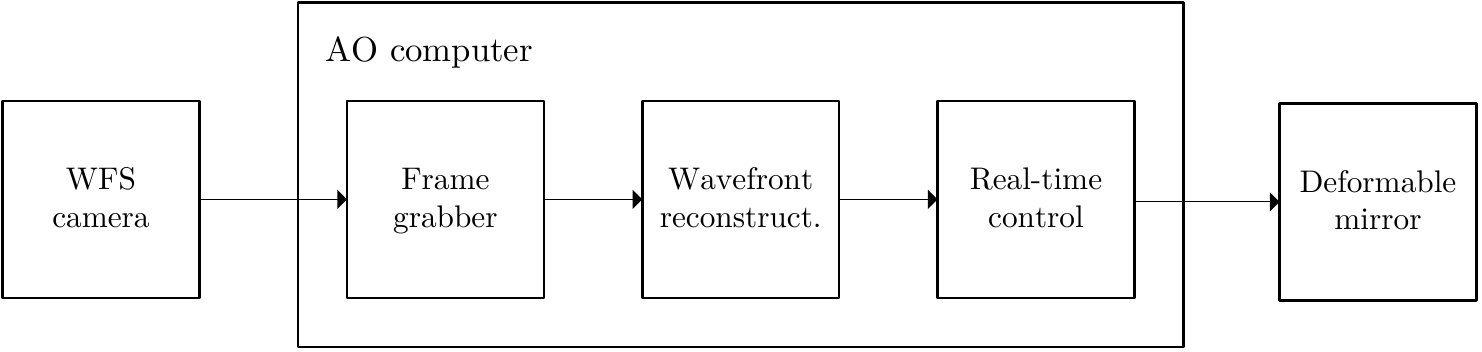}
\caption {AO processing pipeline, consisting in three pieces of software running on the AO computer. The pipeline is triggered on the reception of a frame from the WFS camera, and re-enters an idle state after a new vector of actuations has been sent to the DM. All the elements are required to have a low latency so as to ensure that the actuation vectors sent to the DM correspond to wavefronts that still exist in the real world.}
\label{fig:ao_computer} 
\end{figure*}

In the end, the AO part of AOLI was implemented with three new pieces of software: the frame grabbing software (FG), the WFR and the RTC. The three pieces of software are interconnected in a processing pipeline as shown in Figure~\ref{fig:ao_computer}. This processing pipeline is triggered every time a new frame is produced by the WFS camera (which sets the AO sampling rate), and it re-enters the idle state once a new actuation vector is sent to the DM.

The target environment for all the AO-related software would be a single computer operating with GPUs and under Windows. We selected an Intel Core i7-4790K CPU and an nVidia GeForce GTX Titan Z GPU running the Windows 7 operative system. The following subsections will provide details about each of the three components of the AO processing pipeline.

\subsection {Frame grabbing software}
\label{sec:fg_sw}

The main objective of the FG software is to continuously acquire images from the WFS camera and send them to the WFR software as soon as they are being received, with no further processing in between. Additionally, this software allows configuring the camera parameters and provides real-time information once the acquisition process has started.

Reference AO stars with bright apparent magnitudes may produce images in the EMCCD sensor with unnecessary large signal-to-noise ratios (S/N), with no actual improvement in the accuracy of the reconstructed wavefronts. In those cases, the operator of the instrument would normally decrease the exposure time of the WFS camera so as to get a faster sampling rate, which would enable the RTC software to produce DM actuations more frequently and thus improve the quality of the AO corrections even with bad atmospheric conditions. The upper limit of the sampling rate is set by the camera hardware itself and by the element of the AO processing pipeline whose execution time is longer.

On the other hand, the S/N of the pupil images of faint reference AO stars can indeed be improved by lowering the sampling rate of the camera, but this may have a negative impact on the quality of the AO corrections because the RTC software may not be able to keep up with the speed of variation of the atmospheric turbulence at that specific period of time. In those cases, the recommended solution is to activate the binning function of the EMCCD, which will increase the S/N of the image by combining the charges of adjacent pixels, at the expense of a lower spatial resolution in the acquired pupil images. This would improve the accuracy of the wavefront reconstructions in the cases of low light, although it will not be possible to reconstruct the higher-order modes because of the loss of image resolution.

For an optimal configuration of the camera, it is important to know that its sampling rate does not only depend on the exposure time and the size of the readout region, but also on other parameters such as the frequency of the EMCCD clocks, specially the clock that drives the ADC (analog-to-digital converter). This type of clock fine-tuning has to be done carefully in order not to degrade the S/N.

\subsection{Wavefront reconstruction software}
\label{sec:wfr}

The WFR software takes the two defocused images formed by the TP3-WFS optics and calculates the photon displacements between those two planes by applying the Radon transform \citep{Radon17} over a set of projection angles. The photon displacements are then used to produce an estimation of the slopes of the incident wavefront. Finally, the algorithm outputs a Zernike representation of the reconstructed wavefront \citep{von1934beugungstheorie}, calculated as the least-squares fit between the calculated slopes and the ones that each individual Zernike mode would produce.

The WFR algorithm is clearly the one that has a higher computational cost in the whole AO processing pipeline, so a huge effort was put in optimizing its implementation in order to achieve real-time performance. Among all the possible acceleration methods, it was decided that the WFR would take advantage of GPUs in order to achieve real-time operation, more specifically, by means of the CUDA language. When compared to other FPGA or CPU-based approaches, the GPU implementation was considered a good trade-off from the point of view of development costs, flexibility and re-usability \citep{ao4elt4_31563}.

\begin{figure*}
\centering
\includegraphics[scale=0.9]{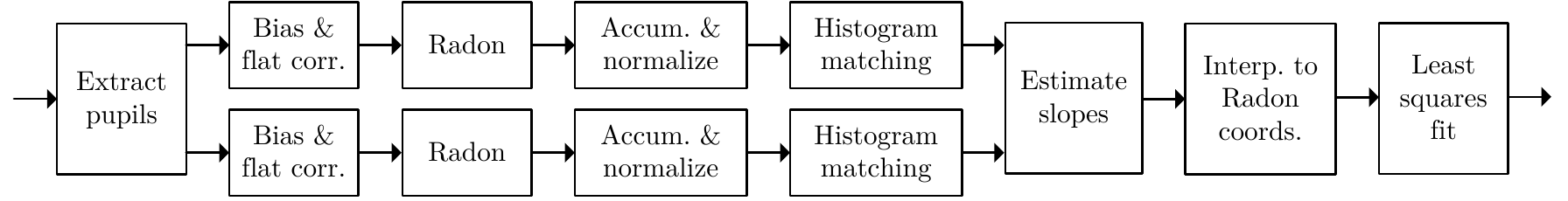}
\caption {WFR software architecture. For each input image, containing two defocused pupil images, a vector of Zernike modes is calculated. The whole algorithm is implemented on GPU, with the exception of the \textit{Extract pupils} block. This is justified because sending full images to the GPU would create a bottleneck in the CPU-GPU communication channel.}
\label{fig:wfr_arch} 
\end{figure*}

Figure~\ref{fig:wfr_arch} provides a graphical summary of the steps of the WFR algorithm, whose equations are further described in \cite{vanDam:02}. Even though the description of the algorithm itself is out of the scope of this paper, the following subsections will give practical considerations about some of the blocks in Figure~\ref{fig:wfr_arch}, which can be helpful for both the development and use of future wavefront sensors using this reconstruction technique, besides the TP3-WFS itself.

\subsubsection{Extract pupils}
\label{sec:pupil_reg}

The first step of the algorithm is to extract the two regions of the input image that correspond to each pupil image. In the case of AOLI, this separation has to be done by software because the optics of the TP3-WFS were designed in such a way that both defocused pupil images fall within the area of a single WFS detector.

In earlier versions of AOLI, each pupil image was sent to one different camera, but in the end this proved to be a bad idea because of two reasons: first, because it was very hard to synchronize the camera hardware and the related software in order to acquire from the two cameras with a high level of synchronism. Synchronism is obviously a requirement because it only makes sense to process pupil that come from the same instant of time. The second reason is that even small differences between the two cameras (quantum efficiency, dynamic range, bias level, noise, etc.) have a considerable impact on the reconstruction quality. Using a single camera is thus an effective way to solve both problems.

\subsubsection{Bias and flat correction}
\label{sec:bias_flat}

The simulations made with different configurations showed that small deviations between the mean bias level of the two pupil regions has a considerable impact on the accuracy of the reconstructed wavefronts. Being an algorithm that bases its calculations on the differences between the two input images, this behaviour was actually expected. Therefore, it is considered mandatory to apply bias corrections as a pre-processing step.

In the same way, it is highly recommended to apply flat-field corrections to the input images, specially to reduce the effect of dust grains on the WFS sensor and optics. During a laboratory test, there was one big particle of dust falling on one pupil image that caused the control loop to enter a peculiar oscillatory regime due to the non-linearity caused by the presence of that particle.

From the point of view of a real-time implementation, it is very convenient to perform the bias and flat-field corrections to each individual pupil image, after they have been extracted to the full image. The rest of the pixels of the original image are not processed anyway, so it is a loss of processing time to correct them as well.

\subsubsection{Radon transform}

One important parameter to configure in the WFR algorithm is the number of projection angles for which the Radon transform will be calculated. This parameter is closely related to the number of Zernike modes to be reconstructed, as the last step of the algorithm is a least-squares adjustment between the measured wavefront slopes and the ones that the different Zernike modes would produce. This means that, for a given number of reconstructed Zernike modes, there is a minimum number of angles that need to be calculated so as to ensure that the least-squares fit is being executed correctly. A larger number of angles means a better ability to represent high-order aberrations in the Radon transforms themselves.

Of course, even though a large number of projection angles would increase the probabilities of performing a correct fit, in practice setting a very high number of angles is a bad idea because the execution time of the algorithm has a linear dependence with the number of angles. For a given number of reconstructed Zernike modes, a trade-off can be found by performing a sweep over the number of angles and determine by inspection whether the result of the static characterization is correct. Section~\ref{sec:sta_res} explains how to analyze the goodness of the static characterization results.

\subsubsection{Least-squares fit}

The final step of the algorithm actually requires to have pre-calculated the mean wavefront slopes that each Zernike mode would produce at perpendicular directions of each projection angle. This operation is particularly expensive from a computational perspective, so it was also GPU-accelerated in spite of not being part of the real-time AO processing pipeline. This enabled us to quickly experiment the effect of selecting different combinations of number of Randon projection angles and Zernike modes.

It is interesting to note that the input to the least-squares fit block in Figure~\ref{fig:wfr_arch} already contains all the information of the reconstructed wavefront, more specifically, the slopes along a set of projection angles. The least-squares fit is just a way of representing that wavefront in a more familiar way, which additionally enables separating some modes of special interest such as tip, tilt and defocus. It remains to be evaluated whether applying the control algorithm to the estimated slopes directly or to their 2-D integral results in a better level of correction.

At the time of writing the present work, the Zernike modes outputted by the TP3-WFS are not expressed in any particular units. While this could be a problem for other applications, it does not have any impact on the RTC control algorithm of AOLI, as this algorithm does not need to know the physical units of the measured wavefronts. Similarly, the results presented in this paper do not lose their validity because they are always intended to be analysed in a differential way (e.g., open loop vs. closed loop).

\begin{figure*}
\centering
\begin{subfigure}{0.2\textwidth}
\centering
\includegraphics[scale=0.8]{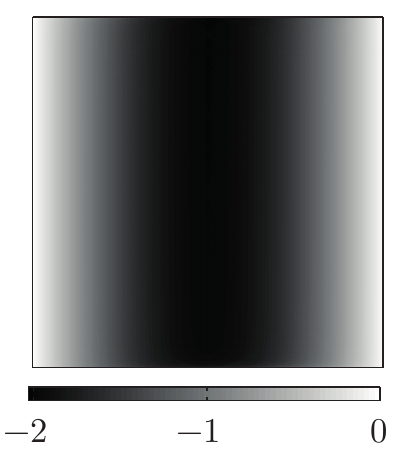}
\caption{Vertical tilt}
\end{subfigure}
\begin{subfigure}{0.2\textwidth}
\centering
\includegraphics[scale=0.8]{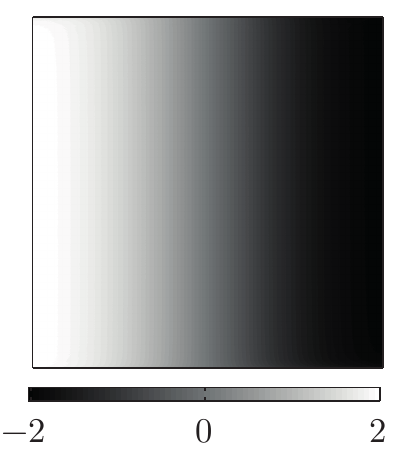}
\caption{Horizontal tilt}
\end{subfigure}
\begin{subfigure}{0.2\textwidth}
\centering
\includegraphics[scale=0.8]{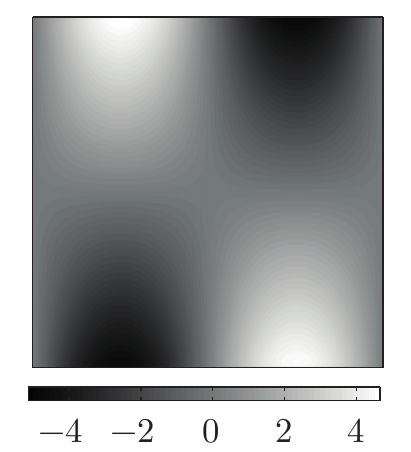}
\caption{Oblique astigmatism}
\end{subfigure}
\begin{subfigure}{0.2\textwidth}
\centering
\includegraphics[scale=0.8]{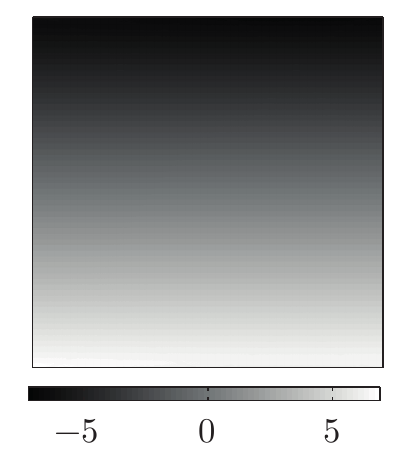}
\caption{Defocus}
\end{subfigure}
\par\bigskip
\begin{subfigure}{0.2\textwidth}
\centering
\includegraphics[scale=0.8]{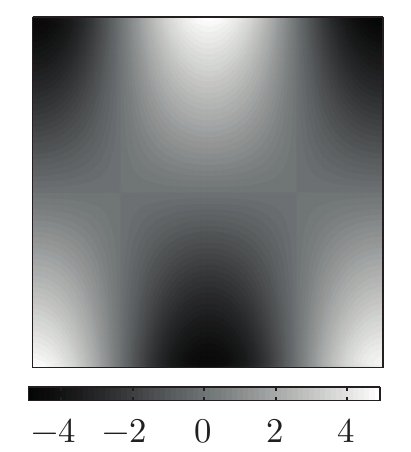}
\caption{Vertical astigmatism}
\end{subfigure}
\begin{subfigure}{0.2\textwidth}
\centering
\includegraphics[scale=0.8]{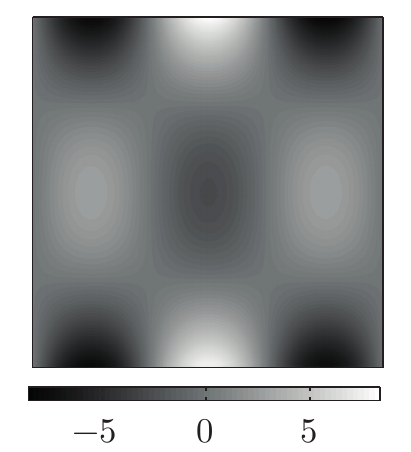}
\caption{Vertical trefoil}
\end{subfigure}
\begin{subfigure}{0.2\textwidth}
\centering
\includegraphics[scale=0.8]{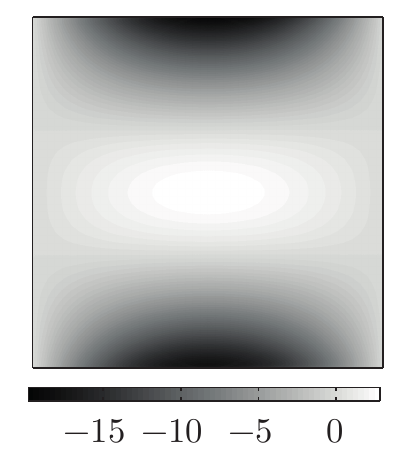}
\caption{Vertical comma}
\end{subfigure}
\begin{subfigure}{0.2\textwidth}
\centering
\includegraphics[scale=0.8]{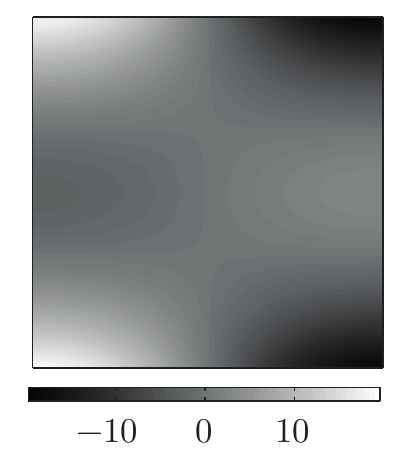}
\caption{Horizontal comma}
\end{subfigure}
\par\smallskip
\caption{Precalculated Zernike slope maps, used during the conversion of the calculated wavefront slopes to vectors of Zernike modes. The \textit{X} axis corresponds to the projection angle $\theta$ ($0 \leq \theta < \pi$), while the \textit{Y} axis corresponds to the Radon coordinate $r$ ($-R \leq r \leq R$, where $R$ is the radius of the aperture).}
\label{fig:wfr_precalc}
\end{figure*}

As a reference for future wavefront sensors willing to implement this reconstruction algorithm, Figure~\ref{fig:wfr_precalc} shows the appearance of the slope maps of the 10 lowest-order Zernike modes (excluding piston, as it cannot be reconstructed).

\subsection{Real-time control software}
\label{sec:rtc_sw}

The RTC software is the one responsible of actuating over the DM in such a way that the effect of atmospheric turbulence on the science camera images is minimized. The control algorithm is based on an array of proportional-integral-derivative (PID) controllers. Besides the real-time functionality itself, the RTC software implements the AO characterization procedures explained later in Section~\ref{sec:sys_char}.

The computations in the RTC software were accelerated by means of the Blaze library \citep{6266939}. The use of this library enabled easy exploiting of the AVX2 instruction set (Advanced Vector Extensions 2) available on the target CPU, which is specially useful for accelerating vector and matrix operations like the ones that are executed in the RTC algorithm. The proper use of these instructions ultimately allowed obtaining better performances than other highly-optimized GPU libraries like cuBLAS, as the latency of the CPU-GPU link constitutes a tight bottleneck for the RTC algorithm.

Figure~\ref{fig:rtc_arch} depicts the architecture of the real-time part of the RTC software. It consists on a processing pipeline that is triggered on reception of a new reconstructed wavefront, and finishes when the new set of actuations is sent to the DM. Below is a brief description of the blocks appearing in Figure~\ref{fig:rtc_arch}:

\begin{figure*}
\centering
\includegraphics[scale=0.9]{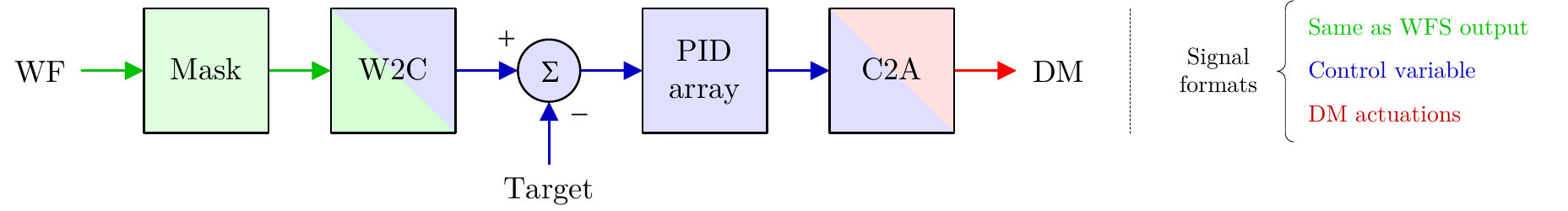}
\caption {RTC software architecture, showing all the configurable blocks and the nature of the signals traveling throughout them.}
\label{fig:rtc_arch} 
\end{figure*} 

\begin{itemize}

\item{\textbf{Mask}: Extracts the elements of interest from the input wavefront. In AOLI it is used to ignore the tip and tilt modes in the AO control loop, as the LI algorithm would remove them anyway in its shift-and-add stage, where the input images are re-centered using a predefined criterion such as the position of the peak pixel.}

\item{\textbf{W2C} (``\textit{wavefront to control}''): Takes the input wavefront and transforms it to a different format just before entering the actual control stage. In AOLI, the W2C block is configured as pass-through.}

\item{\textbf{Target}: Allows setting a control target that is different from 0. This is useful for compensating non-common path between the WFS camera and the science camera (NCPA, Non-Common Path Aberrations).}

\item{\textbf{PID array}: As its name suggests, this block is composed of a set of independent PID controllers, one for each element present at its input vector (actually, an error vector).}

\item{\textbf{C2A} (``\textit{control to actuations}''): transforms the output of the PID array into the specific set of actuations that produce the desired correction. In AOLI, this block is configured to transform Zernike modes into actuation vectors.}

\end{itemize}

\section{System analysis}
\label{sec:sys_char}

In order to have a fully working AO instrument one first needs to obtain information from the control plant (i.e., the elements that are part of the control loop) to configure the parameters of the RTC processing pipeline that were explained in Section~\ref{sec:rtc_sw}. For that purpose, we developed two different procedures, each of them characterizing the AO system under different conditions.

On one hand, the \textit{static characterization} procedure is used to obtain the C2A matrix by inverting the obtained influence matrix. On the other hand, the timing information given by the \textit{dynamic characterization} procedure allows setting proper values to the \textit{P}, \textit{I} and \textit{D} parameters of the PID array. The following subsections will explain these two procedures in detail.

\subsection{Static characterization}
\label{sec:sta}

The static characterization procedure allows obtaining the influence function of the AO system. This function represents the effect of actuating over the DM as seen by the WFS once the DM has reached its steady state.

The influence function is usually represented as the so-called \textit{influence matrix} $\mathbf{I_{M \times A}} = [\mathbf{i}_1,... , \mathbf{i}_A]$, where $M$ is the number of elements at the output of each  WFS sample (normally a set of Zernike modes) and $A$ is the number of actuators of the DM. In this matrix, each row $\mathbf{i}_a$ contains the response of the WFS when a single actuator is pushed with an actuation value equal to 1, whichever the actuation units are. 

From the static point of view, the influence matrix contains all the information that is needed to correct an aberrated input wavefront by actuating over the DM. But that would only possible if the control algorithm knew the vector of actuators that compensates a given input wavefront, which is just the opposite information that the influence matrix provides. Fortunately, it is possible to invert the influence matrix so that it can be used to convert from wavefronts to actuation vectors, even if the matrix is not square. If the matrix is not square, one can apply the Moore-Penrose pseudoinverse \citep{penrose2008} instead of performing a regular matrix inversion.

The acquisition of the influence matrix was performed using the algorithm described in Appendix~\ref{sec:sta_app}. There is a set of parameters that is used to configure the algorithm, namely: $n_{sta}$ (stabilize samples), $n_{pp}$ (push-pull iterations), $n_{avg}$ (average samples) and $a_{val}$ (actuation value).

The parameter $n_{sta}$ sets the amount of time (measured in WFS samples) to wait for the AO system to stabilize after a DM actuation. This does not only depend on the response time of the mirror, but also on the rest of elements on the AO chain. The best way to set a proper value to $n_{sta}$ is to measure the dynamic response of the system (Section~\ref{sec:dyn}) so as to discover the amount of WFS samples that takes for the system to reach a steady state after a step in the input.

One could expect that higher values in $n_{pp}$ and $n_{avg}$ should enable getting measurements with a lower level of noise. However, setting $n_{avg} > 1$ can have a negative effect on the measurement if there is a non-zero mean drift in the input wavefront (either from a star in the sky or from a calibration source). Therefore, it is recommended to set $n_{avg} = 1$ and increase the value of $n_{pp}$ in order to obtain a good signal-to-noise ratio. 

Regarding $a_{val}$, it should be set with a value such that the signal that it produces on the WFS is bigger than the noise, and at the same time ensuring that the AO system (DM, optical path and WFS) does not exit its linear zone during the whole characterization process. The initial position of the DM is not relevant as long as it enables complying with this latter restriction.

The amount of time required to execute the static characterization procedure ($t_{sta}$) depends on the sampling rate of the WFS ($f_{WFS}$), the number of actuators ($A$) and on the following algorithm parameters: $n_{sta}$, $n_{pp} $ and $n_{avg}$. By analyzing the steps of the algorithm, it can be shown that the duration of the static characterization process follows the equation below:

\begin{equation}
\label{eq:sta_time}
t_{sta} = \frac{A \left(n_{sta} + 2 n_{pp} (n_{sta} + n_{avg})\right)}{f_{WFS}}
\end{equation}

\subsection{Dynamic characterization}
\label{sec:dyn}

The dynamic characterization procedure is used to obtain the impulse reponse of the AO system. Unlike the influence function obtained during the static characterization, the impulse response characterizes the system also in the time domain.

The information provided by the impulse response can be useful for designing a proper PID control loop, that is, setting adequate values to the $K_p$, $K_i$ and $K_d$ parameters. Another application of the measured impulse responses is to determine the amount of WFS samples that the AO system needs to reach a stable state after an actuation vector is sent to the DM, which sets a lower limit to the $n_{sta}$ parameter of the static characterization procedure.

The dynamic characterization process consists on two stages. In the first stage, a pre-defined sequence of actuations $d_{in}[j]$ (with $1 \le j \le n_{in})$ is introduced into one actuator of the DM, and the resulting WFS output $\mathbf{d}_{out}[j] = \left [ d_{out,1}[j],...,d_{out,M}[j] \right ]$ (where $M$ is the number of elements at the output of each WFS sample) is recorded as each element of the input sequence is introduced. After that, a post-processing stage is executed in order to obtain the impulse response of the system $\mathbf{h}[j] = \left [h_1[j],...,h_M[j] \right ]$. The specific post-processing that needs to be applied depends on the nature of the selected input sequence. 

\subsubsection{On-line stage}

The steps required to perform the on-line stage of the dynamic characterization over one DM actuator are explained in Appendix~\ref{sec:dyn_app}. There are three configurable parameters in this algorithm: $n_{sta}$ (stabilize repetitions), $n_{avg}$ (average repetitions) and $t_{RTC}$ (duration of each iteration of the RTC algorithm).

It is required that $n_{sta} > 0$ because otherwise the system will not have entered a stationary state by the time the first output sample is read. Setting $n_{avg} > 1$ allows improving the signal-to-noise ratio, though the same effect can be achieved just by increasing the length of the input signal (i.e., increasing $n_{in}$), provided that the selected input signal is not a plain impulse. The parameter $t_{RTC}$ must represent the amount of time that each iteration of the RTC algorithm would take, otherwise there would a discrepancy between the complete AO system and the system which is being measured. The output signal $\mathbf{d}_{out}$ is the one that will be processed during the off-line stage so as to obtain the impulse response $\mathbf{h}[j]$.

As happened in the static characterization, the input parameters of the dynamic characterization have an effect on its execution time $t_{dyn}$, which may be calculated as follows:

\begin{equation}
\label{eq:dyn_time}
t_{dyn} = \frac{n_{in} (n_{sta} + n_{avg})}{f_{WFS}}
\end{equation}

\subsubsection{Off-line stage}

The calculations to be done on the off-line stage depend on the nature of the chosen input sequence. Among the different types of input sequences, we tested three of the most well-known ones: an impulse, white noise and maximum length sequences (MLS) \citep{borish1983efficient, rife1989transfer}. The tests were executed both in computer simulations \citep{2015inac.conf...51C} and with the actual AOLI instrument. The authors finally chose the MLS method because it proved to perform better than the others, meaning that it produced more accurate, less noisy impulse responses with lower measurement durations.

The MLSs were generated with LFSRs (Linear Feedback Shift Registers) using the taps proposed by \cite{ward2012table}. These LFSRs work with the values $-1$ and $1$ instead of $0$ and $1$, thus producing a pseudo-random sequence containing positive and negative pulses. The input sequence is multiplied by a constant $a_{val}$ in order to control the magnitude of the actuation, in a similar way as it was done during the static characterization. The length the MLS sequence $n_{in}$ depends on the MLS order $m$ as indicated in the following equation: $n_{in} = 2^m - 1$.

When the input sequence is a MLS, one can apply the cross-correlation between the MLS input and output in order to get the impulse response of the $i$th element of the WFS output vector (normally a Zernike mode):

\begin{equation}
\label{eq:dyn_mls}
h_i[j] = \frac { \mathcal{F}^{-1} \left ( \mathcal{F} \left ( d_{out,i}[j] \right ) \mathcal{F} \left ( d_{in}[j] \right )^\ast \right ) }{n_{in}(a_{val})^2}
\end{equation}

\section{Results}
\label{sec:results}

The AO system described in the previous sections was extensively tested under different conditions before going to telescope. Finally, on May 2016 the full AOLI instrument was moved to the William Herschel Telescope, were it saw first light on 22nd May 2016. At that night, we managed to close the AO control loop with a natural sky star with the new TP3-WFS, though with limited performance due to unexpected alignment issues. With the lessons learned during that night, further AO-related results were gathered on another commissioning run in October 2016, closing the loop once again but that time with several targets. In this section we will present the AO-related results, obtained both during laboratory and telescope tests.

\subsection{Static characterization}
\label{sec:sta_res}

During laboratory tests, the influence matrix itself proved to be an invaluable tool to learn how to configure the TP3-WFS, which had never been used on an AO system before. Given the fact that the WFR software was designed to also output the measured wavefronts as 2-D surfaces (calculated from the 1-D Zernike vectors), it was easy to determine whether the measurements were being done correctly just by representing each actuator response obtained during the static characterization (Section~\ref{sec:sta}) and comparing it with the expected response.

In a system that works correctly, the 2-D representation of each actuator response should contain a peak that represents the position of the actuator as seen by the WFS. In the case of the TP3-WFS, one just has to ensure that the number of reconstructed Zernike modes is large enough to achieve a resolution that allows identifying such peaks. The specific number of modes that produce such resolution can be obtained either by computer simulations or by performing a sweep over the number of reconstructed modes with the test equipment. In the case of AOLI, the second option was used.

\begin{figure*}
\centering
\includegraphics[scale=1.0]{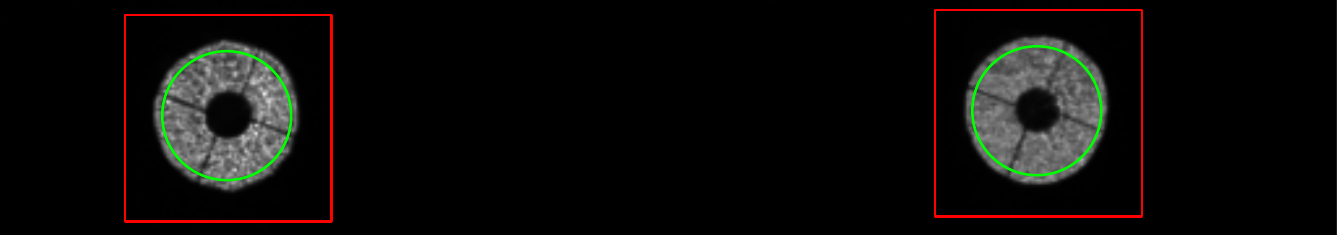}
\caption {Configuration of the TP3-WFS, represented over an actual image read from the WFS camera while pointing to a natural star. The red squares correspond to the extracted pupil regions, while the green circles represent the size of the Zernike functions used for precalculating the slope maps. The background image does not correspond to a full frame but to a limited number of scan lines, in an attempt to maximize the camera sampling rate by reading only the region of interest.}
\label{fig:wfs_conf} 
\end{figure*}

The TP3-WFS was configured as shown in Figure~\ref{fig:wfs_conf}. The processed pupil regions measure 80x80 pixels each, the radius of the Zernike functions used during the precalculations was set to 25 pixels, the number of reconstructed Zernike modes was 153 and the number of Radon angles was 31.  Given that the area of interest is only that of the pupils, the frame grabbing software was configured to read only the 90 scan lines where the pupils were located, in an attempt to maximize the sampling rate.

The rationale for configuring the pupil regions, the Zernike radius and the Radon angles the specified way can be found in Section~\ref{sec:wfr}. Regarding the 153 reconstructed modes, it was tested that closing the loop in laboratory with a lower number of modes led us to a worse PSF (Point Spread Function), while a higher number of modes did not result in a noticeable improvement. Ignoring the fact that LI would eliminate the need of controlling high-order modes, we decided to keep the specified number of modes with the objective of getting the best out of the AO loop alone.

On the other hand, the static characterization parameters were established as follows: $n_{sta} = 2$, $n_{pp} = 25$, $n_{avg} = 1$ and $a_{val} = 0.1$. This combination of parameters produced a good signal-to-noise ratio even on sky tests, with a reasonable execution time. The sampling period of the WFS camera was set to 16.243 milliseconds, and the number of actuators of the DM was $241$. As a result, static characterizations in AOLI took almost 10 minutes, which is coherent with equation (\ref{eq:sta_time}).

\begin{figure*}
\captionsetup[subfigure]{justification=centering}
\centering
\begin{subfigure}{0.2\textwidth}
\centering
\includegraphics[scale=0.7]{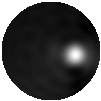}
\caption{Normal actuator,\\ laboratory}
\label{fig:sta_calib_normal} 
\end{subfigure}
\begin{subfigure}{0.2\textwidth}
\centering
\includegraphics[scale=0.7]{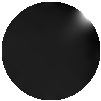}
\caption{Border actuator,\\ laboratory}
\label{fig:sta_calib_border} 
\end{subfigure}
\begin{subfigure}{0.2\textwidth}
\centering
\includegraphics[scale=0.7]{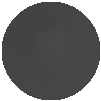}
\caption{Center actuator,\\ laboratory}
\label{fig:sta_calib_center}
\end{subfigure}
\par\medskip
\begin{subfigure}{0.2\textwidth}
\centering
\includegraphics[scale=0.7]{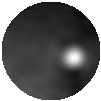}
\caption{Normal actuator,\\ sky star}
\label{fig:sta_star_normal} 
\end{subfigure}
\begin{subfigure}{0.2\textwidth}
\centering
\includegraphics[scale=0.7]{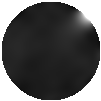}
\caption{Border actuator,\\ sky star}
\label{fig:sta_star_border} 
\end{subfigure}
\begin{subfigure}{0.2\textwidth}
\centering
\includegraphics[scale=0.7]{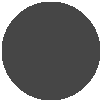}
\caption{Center actuator,\\ sky star}
\label{fig:sta_star_center}
\end{subfigure}
\caption{Static characterization results for a selection of actuators, using either a laboratory calibration source or a natural star as AO reference. The actuators are detected at the expected positions. The center actuator is a special case because it is hidden by the central obscuration of the aperture, making it impossible to detect.}
\label{fig:sta_res}
\end{figure*}

Figure~\ref{fig:sta_res} shows the results of performing the static characterization both in the laboratory (using the instrument calibration source, which simulates the WHT telescope) and with a natural star in the WHT. Instead of representing all the Zernike modes for all the actuators, this figure shows the 2-D surface representation of the modes for a few actuators, in order to ease the interpretation of the result. In the case of the laboratory tests (first row of Figure~\ref{fig:sta_res}) there was no simulated turbulence, while during the measurements with a natural star (second row of Figure~\ref{fig:sta_res}) the atmosphere created a natural turbulence giving a seeing of about 0.9 arcsec.

The results show that the influence functions obtained in the laboratory and in the sky were approximately equal. Of course, the ones obtained in the laboratory were more accurate because there was no turbulence, so they could be used as a reference to estimate the quality of the measurements performed with natural stars.

The 2-D surface representations in Figure~\ref{fig:sta_res} show the peak caused by each actuator in the cases were one would expect a peak. For example, one would expect a peak in Figures~\ref{fig:sta_calib_normal} and \ref{fig:sta_star_normal} because these actuators correspond to the visible area of the WFS (that is, the areas in grey in Figure~\ref{fig:wfs_conf}). In Figures~\ref{fig:sta_calib_border} and \ref{fig:sta_star_border} one can see the response of an actuator located just at the outer border of the Zernike area marked in green in Figure~\ref{fig:wfs_conf}, which can still be detected it because it falls within the pupil region (red square in Figure~\ref{fig:wfs_conf}). Lastly, Figures~\ref{fig:sta_calib_center} and \ref{fig:sta_star_center} show no peaks because the location of this actuator corresponds to the obscured area at the center of the pupils.

The main conclusion of the static characterization tests is that the TP3-WFS is able to reconstruct the high-resolution wavefronts generated by the movement of each individual actuator, even with pupils that have a central obscuration area as in the WHT. Knowing that the pupils in AOLI would have a central obscuration, we prepared an alternative version of the software which calculates annular Zernike modes \citep{Mahajan:81} instead of the regular ones. However, it was not necessary to use it during the commissioning at the telescope because the main software behaved perfectly. 

\begin{figure}
\centering
\includegraphics[scale=0.75]{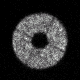}
\quad
\includegraphics[scale=0.75]{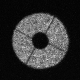}
\caption{A pair of pupil images while pointing at Neptune, extracted from a single camera frame. These images resemble those obtained while pointing to a natural star.}
\label{fig:nep_pupils} 
\end{figure}

\begin{figure}
\centering
\includegraphics[scale=1.4]{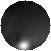}
\caption{Static characterization result of a single visible actuator, using Neptune as the AO reference object. The actuator is detected at the expected position, just as happened when referencing with stars.}
\label{fig:sta_neptune} 
\end{figure}

After verifying that the TP3-WFS apparently produced correct measurements with natural reference stars, it remained to be checked whether the same thing could be said about extended objects such as planets. For this purpose, the telescope was pointed to Neptune (apparent diameter = 2.3 arcsec) and the characterization procedure was executed again. The pupil images acquired by the WFS camera (Figure~\ref{fig:nep_pupils}) were a good initial sign because they looked very similar to the ones that had been obtained while pointing to stars of similar apparent magnitudes. Figure~\ref{fig:sta_neptune} shows the result for a single actuator that is neither in the border nor in the center of the pupil region. The fact that the actuator was correctly detected suggests that the TP3-WFS may allow closing the AO loop with extended objects. This hypothesis will be confirmed later in Section~\ref{sec:closed_loop}.

\subsection{Dynamic characterization}
\label{sec:dyn_res}

The dynamic characterization procedure described in Section~\ref{sec:dyn} was performed on laboratory. The same results can be expected under all conditions, including sky observations.

The parameters of the dynamic characterization were established as follows: $n_{sta} = 2$, $n_{avg} = 10$, $a_{val} = 0.05$ and $t_{RTC} = 0$ seconds. These parameters were set this way because of the reasons already outlined in Section~\ref{sec:dyn}. The simulated delay of the RTC algorithm ($t_{RTC}$) was set to 0 because, in AOLI, this delay is several orders of magnitude lower than the rest of the delays of the system (just a few tens of microseconds). The sampling period of the WFS camera was set to 16.243 milliseconds, and only one actuator was measured: the one at the centre of the DM. The MLS order was set to $m = 9$, so the length of the input sequence was $n_{in} = 511$. As a result of the chosen parameters, dynamic characterizations in AOLI took almost 100 seconds, which is coherent with equation (\ref{eq:dyn_time}).

\begin{figure}
\centering
\includegraphics[scale=0.7]{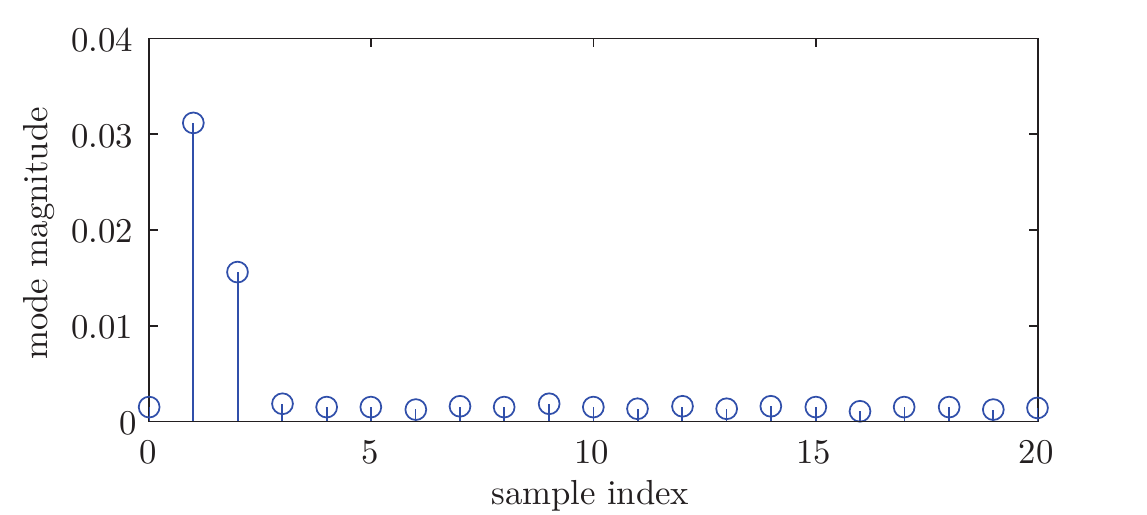}
\caption {Impulse response of the center actuator, defocus mode. The response is distributed over the first two samples. This behaviour can be explained by carefully analyzing the timing of each element in the AO processing pipeline.}
\label{fig:impulse} 
\end{figure}

Figure~\ref{fig:impulse} shows the impulse response of the actuator selected for this test. The impulse response of an actuator is actually a vector of Zernike modes, but in this figure only the defocus mode was represented. This choice is justified by the fact that, for an actuator located at the center of the pupil, only the circular modes (defocus, primary spherical, secondary spherical, etc.) have enough signal level over the noise. From the point of view of the timing analysis, the specific mode that is chosen does not matter as long as it provides enough S/N ratio. For other purposes different than this analysis, one could measure the response of any actuator over any mode by executing the dynamic characterization over a long enough time, but the shape of the resulting impulse response would be the same.

\begin{figure*}
\centering
\includegraphics[scale=0.8]{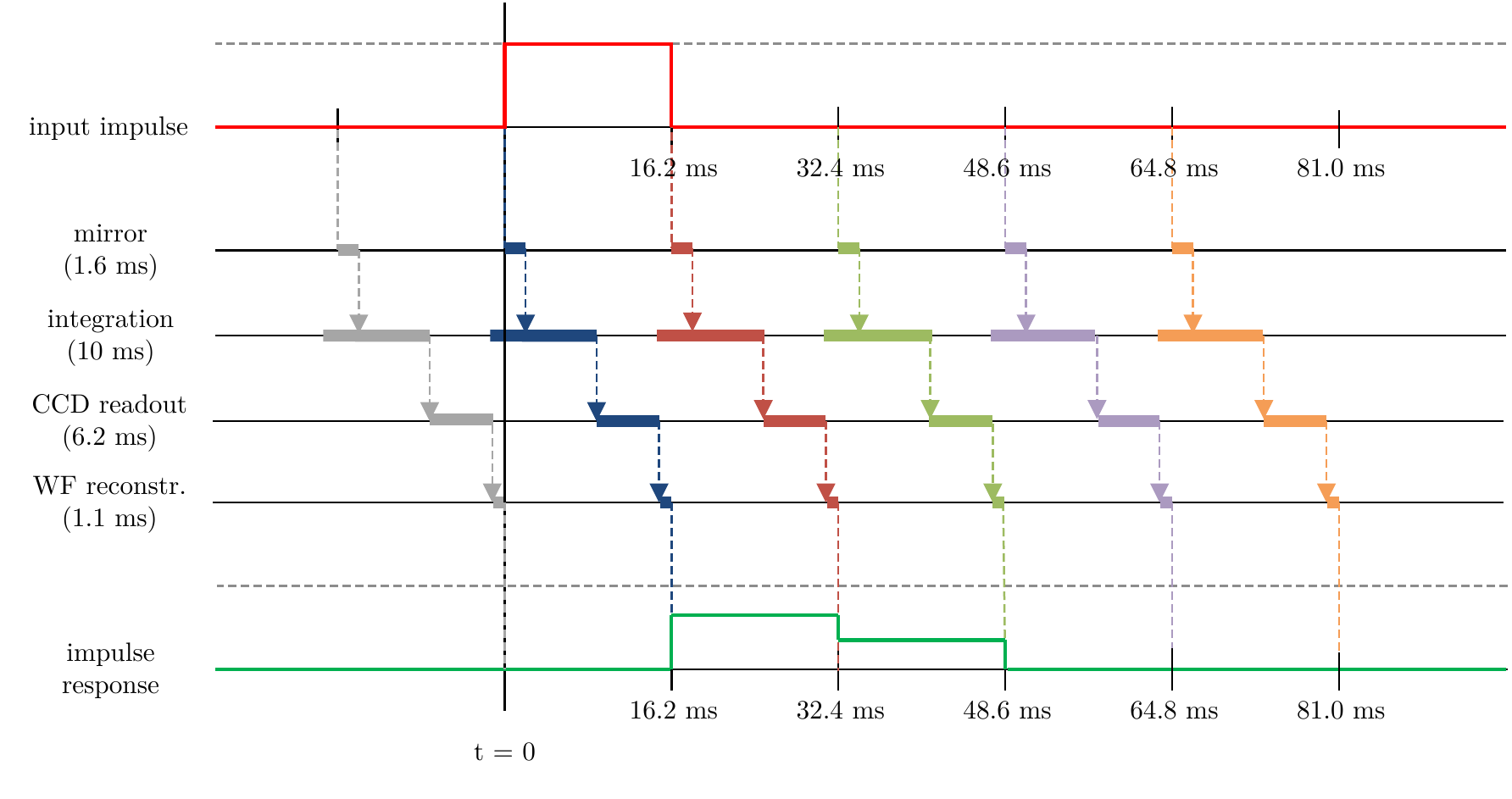}
\caption {Timing diagram of the insertion of an impulse into the AO system, which explains the obtained impulse response. Only the top four delaying elements (mirror response time, integration period, CCD readout time and wavefront reconstruction latency) have been depicted in this diagram. Other elements such as the communication latency between the programs of the AO processing pipeline have not been drawn because they are not significant when compared to the rest of delays. In the diagram, each color in the intermediate lines corresponds the events that are triggered with the arrival of each sample (i.e., an image from the WFS camera).}
\label{fig:timing_diag} 
\end{figure*}

In Figure~\ref{fig:impulse}, the first feature that attracts the attention is the fact that the impulse response only has signal in two of the samples, the rest of the samples only contain noise. The rationale behind this behaviour is hard to realize unless one has some extra information about the timing of the system. That is exactly what Figure~\ref{fig:timing_diag} tries to illustrate. This figure represents a timeline of the tasks that occur during the dynamic characterization of the system, using actual timing data from AOLI.

The first conclusion that can be extracted from Figure~\ref{fig:timing_diag} is that transients of the DM cannot be observed by the WFS cameras. Given the fact that the integration period is much longer than the settling time of the DM, transients are somehow averaged during the integration process, and the camera sees approximately the same signal that it would see with a perfectly steep input. That is, from the point of view of the WFS, the DM responds instantly.

Further inspection of Figure~\ref{fig:timing_diag} reveals that the reason why the response to an impulse is divided in two samples is related to the instant of time along the integration period in which the actuation occurs, which in turn depends on the rest of delays of the system. For a system with timings similar to those of AOLI, it makes sense that the impulse response has two samples different from zero at most. The actual position and relative amplitude of these samples only depend on the accumulated delay of the tasks that are part of the plant of the system under control.

\subsection{Closed loop operation on the telescope}
\label{sec:closed_loop}

After gathering the required information and expertise to close the control loop with the TP3-WFS on laboratory tests, we managed to successfully close the control loop with a natural star in the first commissioning night of the complete AOLI instrument in May 2016, in spite of the bad atmospheric conditions (between 1 and 2 arcsec) and an unexpected alignment problem that was discovered just as the telescope was pointed to the first star. The effects of this alignment problem, located at the instrument-telescope interface, accumulated during the night and made it impossible to close the loop reliably with other targets. In addition, the conclusions that could be extracted from the first target were limited because the acquisition parameters of the science camera had been accidentally set in such a way that the detector got saturated on closed loop operation, causing the so-called blooming effect.

The following commissioning run took place in October 2016. This time a lot of effort was put on ensuring a good alignment between the optical axes of the instrument and the telescope. One of the methods used to align the AO subsystem with the telescope during daytime was to open its petals, turn on the interior dome lights, and then perform an iterative re-adjustment of the instrument so that both the telescope and the calibration unit produced the defocused pupil images at the same positions of the WFS detector, regardless the defocus distances.

Despite the improved alignment, it was determined that static characterizations performed with the instrument calibration unit can degrade the control performance when used with natural stars. The reason is that the slightest discrepancies between the optical axis at the output of the calibration unit and at the output of the telescope make the incoming light fall within different areas of the DM surface and the WFS detector, thus changing the static response of the AO system. Therefore, it is highly recommended to be cautious about this and spend some minutes calibrating with a natural star at least once at the beginning of the night. The calibration process can be accelerated by choosing a bright reference star, which allows setting a fast AO sampling rate and still obtain a good S/N.

During the telescope tests, the TP3-WFS was configured as explained in Section~\ref{sec:sta_res}. Depending on the magnitude of the AO reference object, the exposure time of the WFS camera and the number of scan lines was tuned to set a proper AO sampling rate, following the guidelines given in Section~\ref{sec:fg_sw}. Regarding the configuration of the control loop, most of the tests were performed discarding the tip-tilt out of the 153 measured modes. The rationale behind this decision can be found in Section~\ref{sec:rtc_sw}. Even though the PID parameters had already been tuned in the laboratory using the information obtained during the dynamic characterization, the PID parameters were further tuned using the real turbulence from the sky, seeking a fast and stable response. The PID parameters were finally set as follows: $0.4 \leq K_p \leq 0.8$ (depending on the atmospheric conditions), $K_i = 0.4 * f_{WFS}$ and $K_d = 0$ for all the 151 controlled modes.

\begin{figure}
\centering
\includegraphics[scale=0.7]{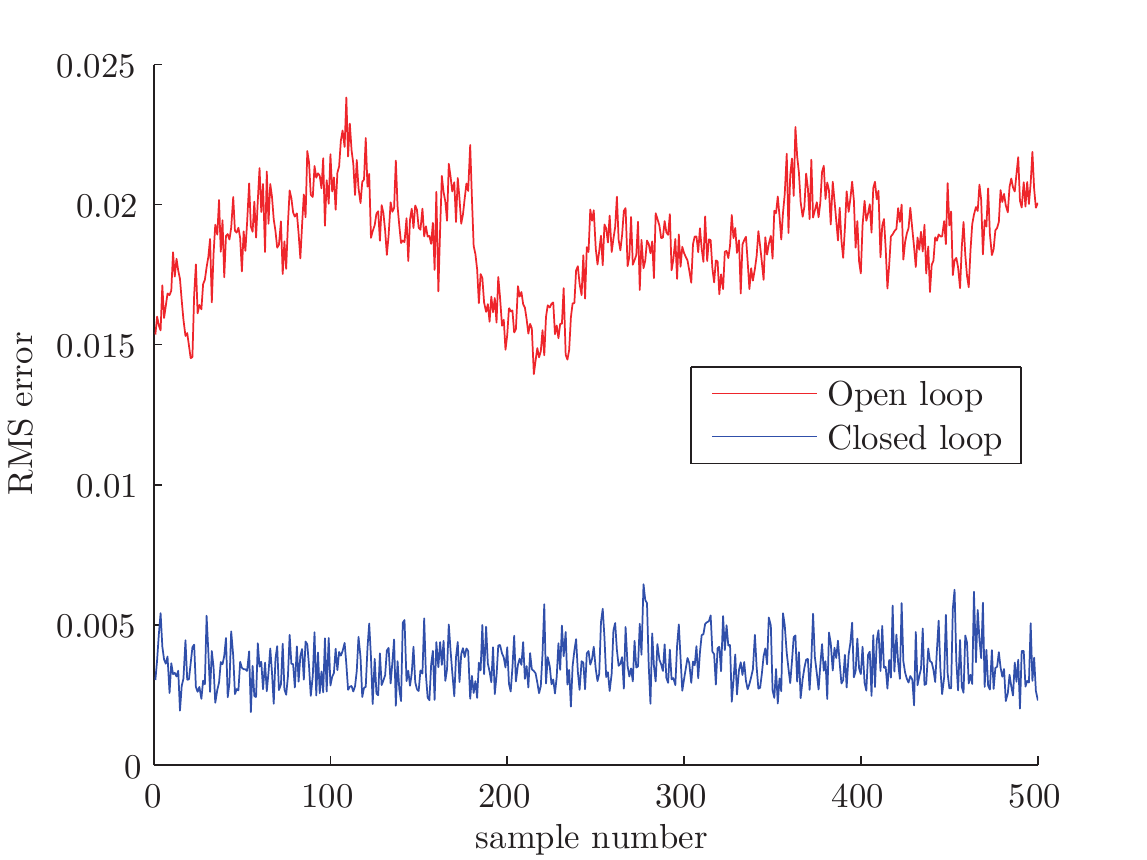}
\caption{RMS error of measured wavefront along time, using HD207470 as AO reference. Closing the control loop reduces both the average RMS error and its standard deviation.}
\label{fig:rms_HD207470}
\end{figure}

Figure~\ref{fig:rms_HD207470} shows the root-mean-squared (RMS) plots of the error of the reconstructed wavefronts when using HD207470 (I magnitude = 7.5) as AO reference star. The error is calculated with respect to the control target, always set to a $0$-vector of Zernike modes with the exception of a manually-tuned defocus term that compensates the NCPA between the WFS camera and the science camera.  When comparing open loop and closed loop operation, Figure~\ref{fig:rms_HD207470} shows an improvement factor of 5.23 in the average RMS, as well as an improvement factor of 2.03 in the standard deviation, which constitutes clear evidence that the control loop is operating correctly.

\begin{figure}
\centering
\includegraphics[scale=0.7]{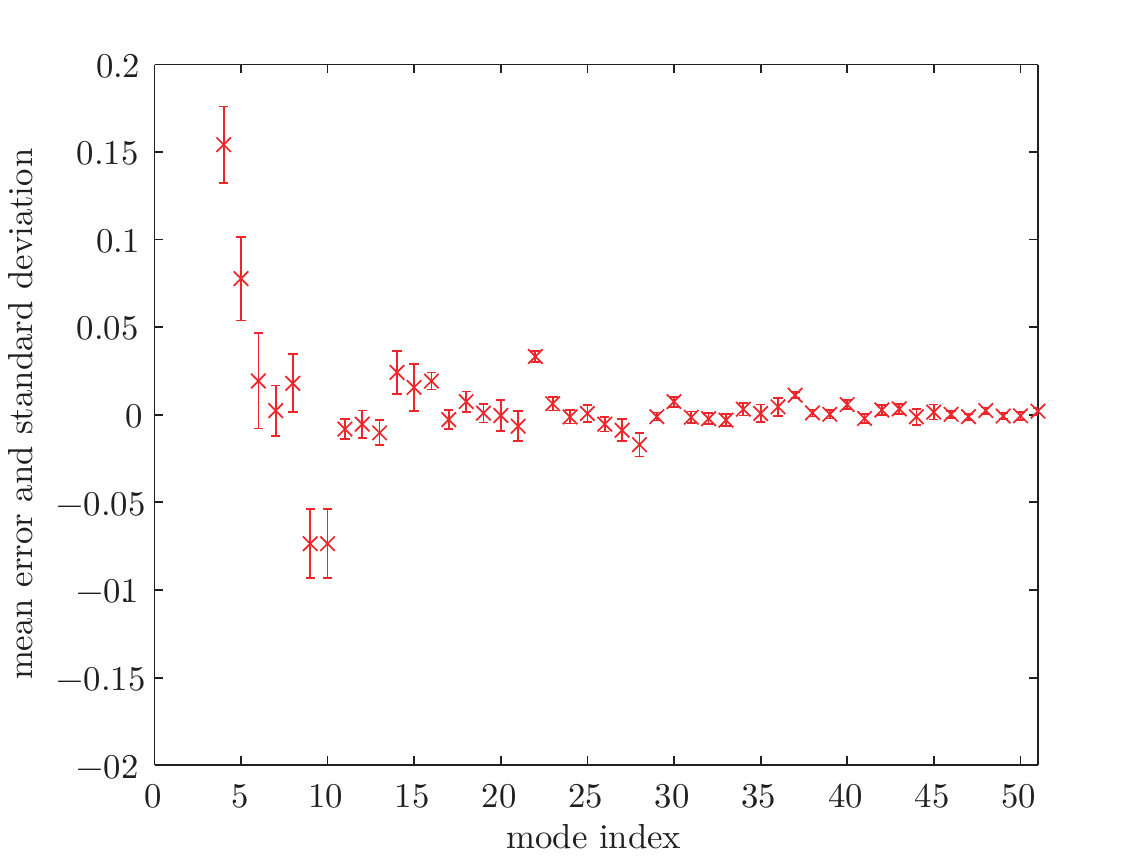}
\caption{Mode-by-mode analysis of HD207470, open loop.}
\label{fig:modes_ol}
\end{figure}

A more in-depth analysis was performed over the same AO reference star, obtaining statistics for each measured Zernike mode. The results are shown in Figures~\ref{fig:modes_ol}. These figures represent the average values and the standard deviation of the error in the first 50 modes (excluding tip and tilt) in open loop and closed loop operation, respectively. The modes have been ordered using the Noll notation \citep{noll1976zernike}. Once again, these plots confirm the correct operation of the control loop, as the average value and standard deviation of each mode is reduced due to the closing of the loop. It was also checked that modifying a mode in the control target vector indeed changed the average value measured for that mode in the specified amount.

\begin{figure}
\centering
\includegraphics[scale=0.7]{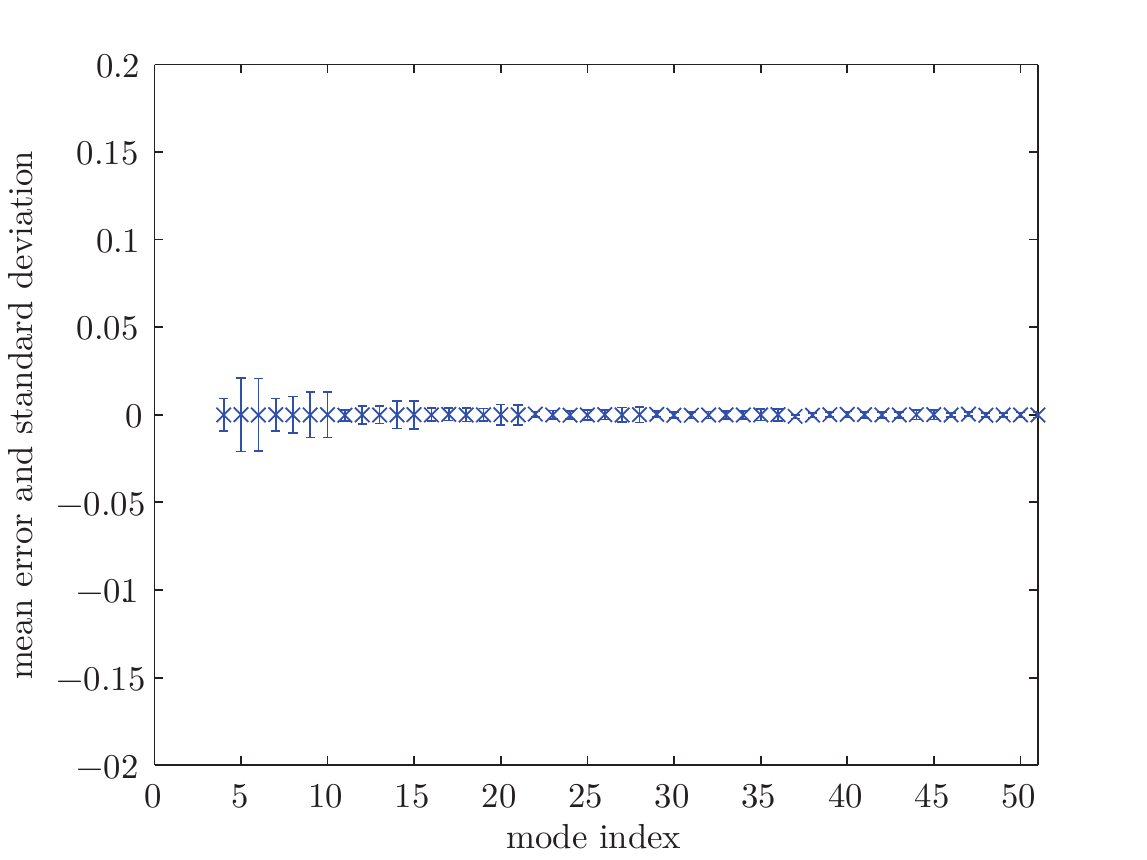}
\caption{Mode-by-mode analysis of HD207470, closed loop.}
\label{fig:modes_cl}
\end{figure}

Similar results were obtained from other AO reference targets different from HD207470. Such results are summarized in Table~\ref{tab:rtc_perf}. The RMS error plots for each target are analogous to the ones presented in Figure~\ref{fig:rms_HD207470}, but modulated by the different RMS averages and standard deviations. For example, the closed-loop error plot for HIP10644 would be shifted down with respect to the plot for HD207470, and it will be less noisy. On the other hand, the lower noise would result in shorter error bars in the mode-by-mode analysis.

\begin{table}
\small
\centering
\begin{tabular}{lccccc}
\toprule
AO ref. & Mag. & $t_{WFS}$ & RMS avg. & RMS std. 	\\
\midrule
HIP10644 & $4.2$ (\textit{I})	& $15.2$ ms	& $1.35 \times 10 ^{-3}$ & $2.16 \times 10 ^{-4}$\\
HDS389AB & $\approx$ $5.5$ (\textit{I})	& $15.2$ ms	& $1.71 \times 10 ^{-3}$ & $3.84 \times 10 ^{-4}$\\
HD207470 & $\approx$ $7.5$ (\textit{I})	& $21.5$ ms	& $3.57 \times 10 ^{-3}$ & $8.46 \times 10 ^{-4}$\\
Neptune & 7.7 (int.)	& $15.2$ ms	& $2.93 \times 10 ^{-3}$ & $7.09 \times 10 ^{-4}$\\
\bottomrule
\end{tabular}
\caption{Control performances for some AO reference objects. The table expresses the dependence between the magnitude of the reference object and the resulting control performance.}
\label{tab:rtc_perf}
\end{table}

One conclusion of Table~\ref{tab:rtc_perf} is that the performance of the control loop depends of the magnitude of the reference object, just as expected from any AO system. Unfortunately, due to a hardware failure that was discovered after the last commissioning run, it was not possible to perform an empirical study about the faintest magnitude of the reference star with which the AO subsystem can operate reliably. One of the EMI (electromagnetic interference) filters of the acquisition card of the WFS camera was physically damaged, resulting in a considerable degradation of the sensibility of the camera. The field study about the performance of the control loop under extremely low light levels is therefore postponed for future commissioning nights where this problem will be hopefully fixed.

\begin{figure}
\centering
\includegraphics[scale=0.7]{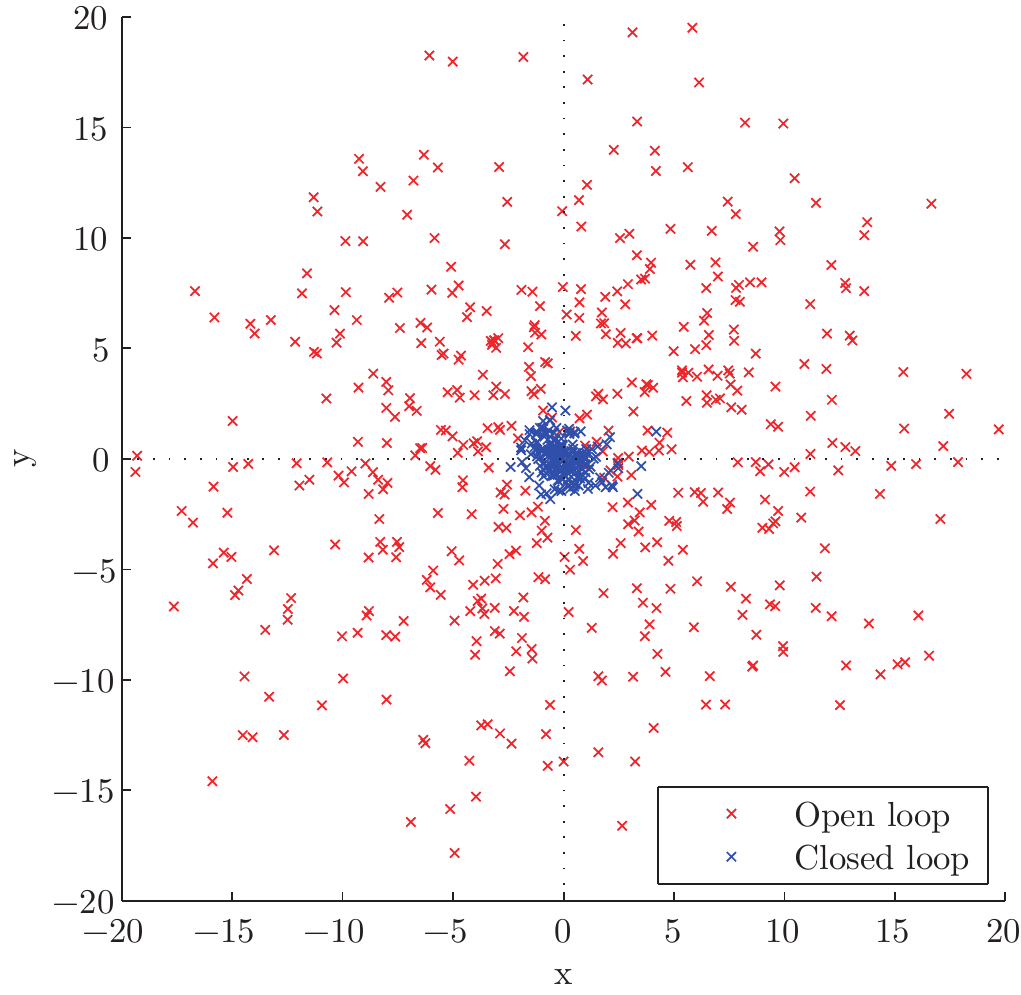}
\caption{Coordinates of Neptune over the science camera. The dispersion of such coordinates is clearly reduced when applying AO with tip-tilt control activated.}
\label{fig:nep_coords}
\end{figure}

Even though AOLI was not originally intended to do science with extended objects, Table~\ref{tab:rtc_perf} also presents the obtained control performances while pointing to Neptune, which confirm the ability of the TP3-WFS to work with extended objects. Further confirmation is presented in Figure~\ref{fig:nep_coords}, which plots the coordinates of Neptune on the science camera when the tip and tilt modes are controlled along with the rest of the modes. This figure is a clear evidence that the tip and tilt modes are being controlled correctly: the standard deviation of the euclidean distance to the mean position is 5.02 pixels in open loop operation, while in closed loop it is only 0.66. Regarding the rest of the modes, due of the lack of features in the surface of Neptune, the only possible test was to adjust the control target of the defocus mode and visually check how it indeed had a proportional effect on the blurriness of the science image.

The results presented so far demonstrate the suitability of the TP3-WFS as the input of an adaptive optics control loop, which was the main purpose of this paper. Nevertheless, further preliminary results will be presented in the following paragraphs regarding the improvement obtained in the images acquired by the science camera. A more comprehensive analysis of the results at the science segment of AOLI, including the application of the lucky imaging algorithm, is scheduled as future work.

\begin{figure*}
\centering
\begin{subfigure}{0.24\textwidth}
\centering
\includegraphics[scale=0.8]{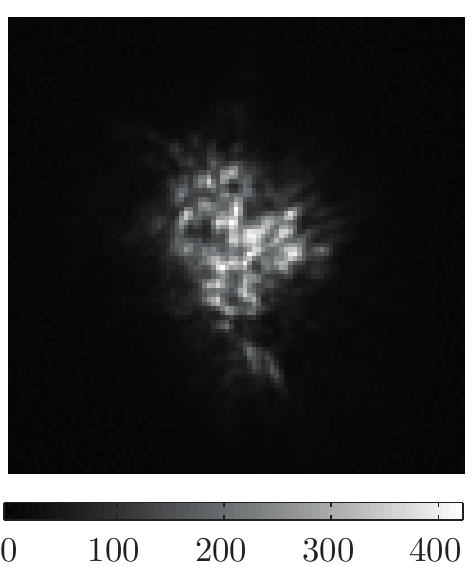}
\caption{Open loop (worst)}
\label{fig:worst_ol} 
\end{subfigure}
\begin{subfigure}{0.24\textwidth}
\centering
\includegraphics[scale=0.8]{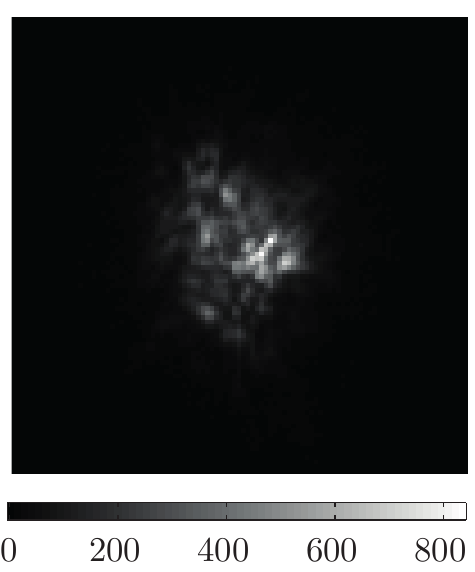}
\caption{Open loop (typical)}
\label{fig:typ_ol} 
\end{subfigure}
\begin{subfigure}{0.24\textwidth}
\centering
\includegraphics[scale=0.8]{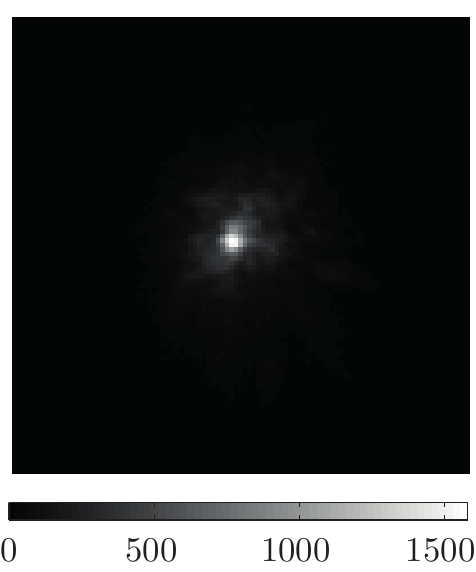}
\caption{Closed loop (typical)}
\label{fig:typ_cl}
\end{subfigure}
\begin{subfigure}{0.24\textwidth}
\centering
\includegraphics[scale=0.8]{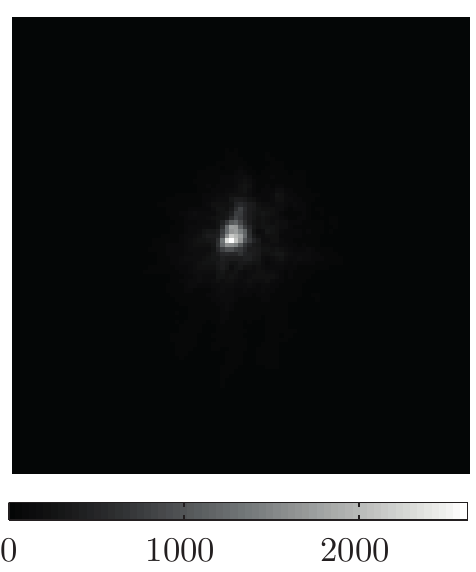}
\caption{Closed loop (best)}
\label{fig:best_cl} 
\end{subfigure}
\caption{Images of HIP10644 with and without AO corrections, showing worst, typical and best images according to the maximum pixel value. It seems clear that the closure of the control loop has a positive impact on the quality of the acquired images.}
\label{fig:control_sci}
\end{figure*}

Figure~\ref{fig:control_sci} shows a set of science images of HIP10644, in open and closed loop, all acquired with 30 ms exposure time. By looking at the speckles, it is clear that the RTC algorithm is effectively reducing the effect of atmospheric aberrations on the science image. The average value of the maximum pixel of each image is 846 in the case of open loop operation, while in closed loop the average value is 1577. This means that the average of the improvement factor of the Strehl ratio is 1.864.

\begin{figure}
\centering
\includegraphics[scale=0.7]{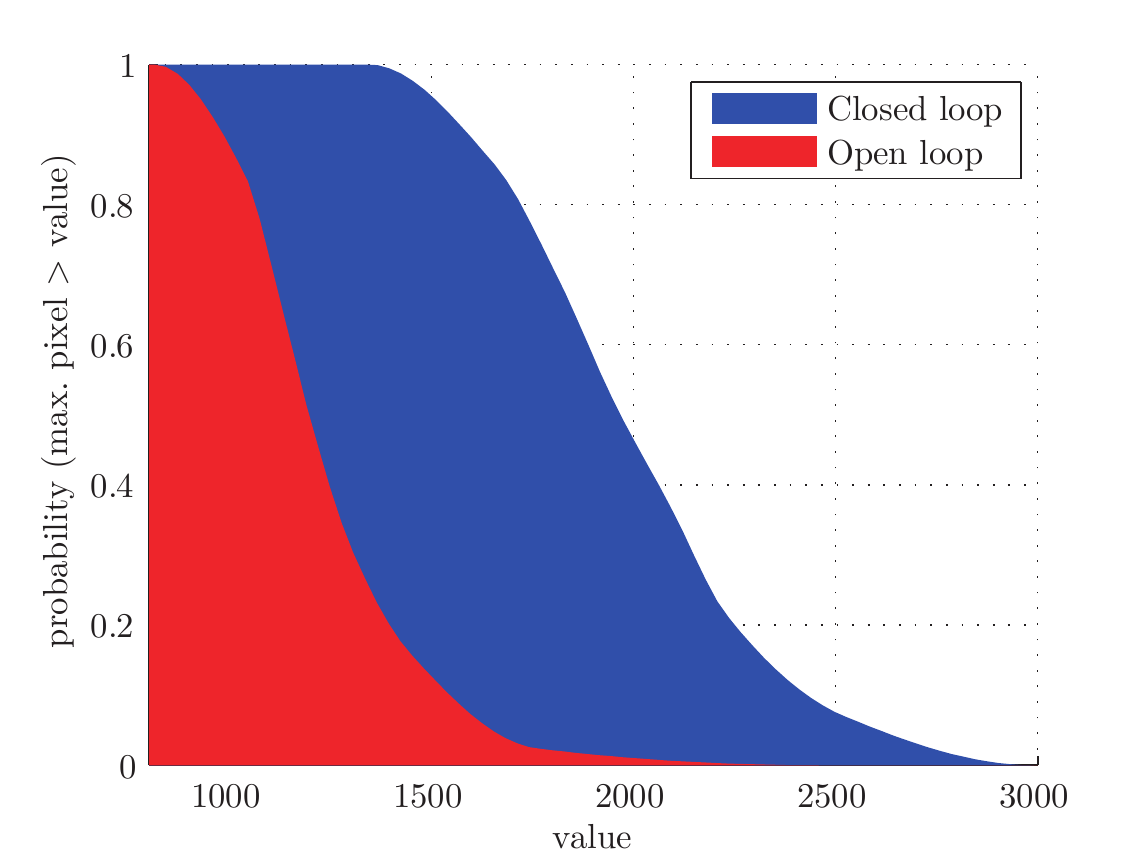}
\caption{Probability of obtaining an image whose maximum pixel value is higher than a given value, calculated from HIP10644. According to this plot, closing the AO loop gives a higher probability of getting an image of a given quality.}
\label{fig:cdf}
\end{figure}

Finally, Figure~\ref{fig:cdf} plots the probability of obtaining an image whose peak pixel value is higher than a given value. The probabilities have been estimated from the images of HIP10644. This figure is specially interesting from the point of view of lucky imaging because it somehow represents the odds of getting a lucky image. For example, for a given number of images, the percentage of images whose maximum pixel value is over 2000 would be around 1\% in open loop, while in closed loop it would be as high as 46.5\%. This could mean an improvement from the point of view of the LI algorithm, which may be able to produce similar results with a significantly lower amount of processed images. This hypothesis, among others, will be thoroughly tested in future works that focus on the science segment of AOLI.

\section{Conclusions}
\label{sec:conclusions}

This paper has presented the first use of the TP3-WFS in a real-time adaptive optics system, with successful results. The analyses, practical considerations and results presented in this paper thus pave the way for new developments besides AOLI willing to use this new type of wavefront sensor, which theoretically provides better sky coverage than previously existing wavefront sensors. The empirical study needed to confirm this latest statement is scheduled for future commissioning runs.

The closure of the control loop improved the average RMS of the 151 controlled modes in a factor of 5.23, which resulted in a clear improvement in the quality of the acquired speckle images, whose average Strehl ratio was increased by a factor of 1.864. In addition, it was confirmed that the reconstruction technique used by the TP3-WFS is able to work with extended objects such as planets, even though such targets fall out of the initial purpose of AOLI.

The presented AO-only results are encouraging due to the fact that they represent just a fraction of what can be achieved by the instrument with the combination of the AO+LI parts. In this regard, future works will focus on thoroughly analyzing the LI performance (open loop vs. closed loop). The preliminary analyses presented in this paper point to a significant improvement of the percentage of lucky images, anticipating a positive outcome also on the science segment.

Regarding the future tests that will be executed to measure the faintest magnitude with which the AO control loop can be closed reliably, two main strategies will be used to maximize the sensibility of the TP3-WFS. First, the radius of the pupil images over the detector will be reduced so as to concentrate better the incoming flux. As explained previously in the article, this would improve the S/N ratio at expense of a reduction in the spatial resolution of the sensed wavefront. The loss of resolution will not pose a problem for AOLI because it was calculated that compensating just the lowest-order modes would enable the LI algorithm to work as expected. The second strategy for maximizing the sensibility would be to use the EMCCDs in photon counting mode instead of the regular imaging mode that was used during the tests described in this article, thus matching the worst case scenario that was assumed in computer simulations. We expect to reach the required limiting magnitude for AO (mag. 16 in the \textit{I} band) by the combination of these two ideas.

\section*{Acknowledgements}
This work was supported by the Spanish Ministry of Economy under the projects AYA2011-29024, ESP2014-56869-C2-2-P, ESP2015-69020-C2-2-R and DPI2015-66458-C2-2-R, by project 15345/PI/10 from the Fundaci\'on S\'eneca, by the Spanish Ministry of Education under the grant FPU12/05573, by project ST/K002368/1 from the Science and Technology Facilities Council and by ERDF funds from the European Commission. The results presented in this paper are based on observations made with the William Herschel Telescope operated on the island of La Palma by the Isaac Newton Group in the Spanish Observatorio del Roque de los Muchachos of the Instituto de Astrof\'isica de Canarias. Special thanks go to Lara Monteagudo and Marcos Pellejero for their timely contributions.




\bibliographystyle{mnras}
\bibliography{aoli_control}



\appendix

\section{Static characterization algorithm}
\label{sec:sta_app}

The static characterization algorithm described in Algorithm~ \ref{alg:sta} \textit{pushes} and \textit{pulls} each actuator with respect to a reference position in order to calculate the slope of the function that relates the WFS readout with the movement of that actuator. The push-pull technique is applied so as to avoid hysteresis effects to affect the measurement. Additionally, all measurements are averaged a given number of times in order to improve the S/N ratio.

\begin{algorithm}[H]
\caption{Static characterization.}
\label{alg:sta}
\begin{algorithmic}
\REQUIRE {$n_{pp} > 0$ and $n_{avg} > 0$ and $a_{val} > 0$}
\STATE $\mathbf{I} \leftarrow \mathbf{0}$.
\FOR {$a = 1$ to $A$}
	\STATE Set DM to initial position.
	\STATE Discard $n_{sta}$ samples from the WFS.
	\FOR {$i = 1$ to $n_{pp}$}
		\STATE Using the DM initial position as a reference, push the $a$th actuator the amount set by $a_{val}$ (i.e., add $a_{val}$).
		\STATE Discard $n_{sta}$ samples from the WFS.
		\STATE Accumulate $n_{avg}$ samples from the WFS.
		\STATE Add the accumulated result into $\mathbf{i}_a$.
		\STATE Using the DM initial position as a reference, pull the $a$th actuator the amount set by $a_{val}$ (i.e., subtract $a_{val}$).
		\STATE Discard $n_{sta}$ samples from the WFS.
		\STATE Accumulate $n_{avg}$ samples from the WFS.
		\STATE Subtract the accumulated result from $\mathbf{i}_a$.
	\ENDFOR
\ENDFOR
\STATE $\mathbf{I} \leftarrow \mathbf{I} / (2 \times n_{avg} \times n_{pp} \times a_{val})$.
\end{algorithmic}
\end{algorithm}

\section{Dynamic characterization algorithm.}
\label{sec:dyn_app}

The dynamic characterization algorithm described in Algorithm~ \ref{alg:dyn} allows obtaining the impulse response of the desired actuator. In a first stage, the system is put in a steady state by introducing the input sequence at least once. Then, the same input sequence is introduced a given number of times, but also recording the output of the system every time a new sample is introduced, ultimately producing an averaged output sequence.

\begin{algorithm}[H]
\caption{Dynamic characterization}
\label{alg:dyn}
\begin{algorithmic}
\REQUIRE {$n_{sta} > 0$, $n_{avg} > 0$ and $t_{RTC} < 1 / f_{WFS}$}
\STATE $\mathbf{d}_{out}[j] \leftarrow \mathbf{0}, \forall j$
\FOR {$i = 1$ to $n_{sta}$} 
	\FOR {$j = 1$ to $n_{in}$}
		\STATE Discard a sample from the WFS.
		\STATE Wait $t_{RTC}$.
		\STATE Using the DM initial position as a reference, push the selected actuator the amount set by $d_{in}[j]$ (i.e., add $d_{in}[j]$).
	\ENDFOR
\ENDFOR
\FOR {$i = 1$ to $n_{avg}$}
	\FOR {$j = 1$ to $n_{in}$}
		\STATE Read a sample from the WFS.
		\STATE Accumulate the new sample into $\mathbf{d}_{out}[j]$.
		\STATE Wait $t_{RTC}$.
		\STATE Using the DM initial position as a reference, push the selected actuator the amount set by $d_{in}[j]$ (i.e., add $d_{in}[j]$).
	\ENDFOR
\ENDFOR
\STATE $\mathbf{d}_{out}[j] \leftarrow \mathbf{d}_{out}[j] / n_{avg}, \forall j$
\end{algorithmic}
\end{algorithm}


\bsp	
\label{lastpage}
\end{document}